\documentclass[preprints,article,accept,moreauthors,10pt,a4paper]{mdpi}
\usepackage{amsfonts,amssymb,amsmath,amsthm,txfonts}
\usepackage{graphicx,epsfig}
\usepackage{epstopdf}
\usepackage{subfigure}
\usepackage[normalem]{ulem}
\usepackage{bbm,bm}
\usepackage[utf8]{inputenc}
\usepackage{hyperref}
\usepackage[linesnumbered,ruled,vlined]{algorithm2e}
\usepackage{algpseudocode}

\newcommand{\ket}[1]{\mbox{$ | #1 \rangle $}}
\newcommand{\bra}[1]{\mbox{$ \langle #1 | $}}

\newcommand{\tr}{\mathrm{tr}}

\newcommand{\cF}{\mathcal{F}}

\firstpage{1}
\pubvolume{xx}
\issuenum{1}
\articlenumber{1}
\pubyear{2018}
\copyrightyear{2018}
\history{}
\Title{Reliable optimization of arbitrary functions over quantum measurements}

\Author{Jing Luo and Jiangwei Shang*}

\AuthorNames{Jing Luo and Jiangwei Shang*}

\address[1]{%
Key Laboratory of Advanced Optoelectronic Quantum Architecture and Measurement of
Ministry of Education, School of Physics, Beijing Institute of Technology, Beijing 100081, China}

\corres{\hangafter=1 \hangindent=1.05em \hspace{-0.82em} Correspondence: jiangwei.shang@bit.edu.cn}

\date{\today}

\abstract{
As the connection between classical and quantum worlds, quantum measurements play a unique role in the era of quantum information processing.
Given an arbitrary function of quantum measurements, how to obtain its optimal value is often considered as a basic yet important problem in various applications.
Typical examples include but not limited to optimizing the likelihood functions in quantum measurement tomography, searching the Bell parameters in Bell-test experiments, and calculating the capacities of quantum channels. 
In this work, we propose reliable algorithms for optimizing arbitrary functions over the space of quantum measurements by combining the so-called Gilbert's algorithm for convex optimization with certain gradient algorithms. 
With extensive applications, we demonstrate the efficacy of our algorithms with both convex and nonconvex functions.
}


\begin{document}
\section{Introduction}\label{Sec:Introduction}
In quantum information science, numerous complex mathematical problems remain to be solved. Since the set of quantum states as well as quantum measurements form convex sets, various important tasks in this field, such as the calculation of ground state energy, violation of the Bell inequality, and the detection and quantification of quantum entanglement \cite{guhne2009entanglement,amico2008entanglement}, conform to the framework of convex optimization theory.
The primary tool in convex optimization is semidefinite programming (SDP) \cite{grant2008graph,Boyd2006ConvexO}, which can be used to derive relaxed constraints and provide accurate solutions for a large number of computationally challenging tasks.
However, serious drawbacks also exist for SDP including its slow computation speed and low accuracy. For instance, SDP can only compute up to four qubits in quantum state tomography (QST), while improved superfast algorithms \cite{shang2017superfast} can quickly go up to eleven qubits with a higher precision.
Consequently, developing more efficient algorithms in convex optimization is becoming more and more crucial as quantum technologies rapidly advance.

Recently, an efficient convex optimization algorithm \cite{brierley2016convex} was proposed by Brierley \emph{et al.} based on the so-called Gilbert's algorithm \cite{gilbert1966iterative}.
Concurrently, Ref.~\cite{montina2016can} used Gilbert's algorithm to investigate whether nonlocal relationships can be distinguished in polynomial time. 
In Ref.~\cite{shang2018convex}, Gilbert's algorithm was employed as a tool to satisfy certain constraints, based on which two reliable convex optimization schemes over the quantum state space were proposed. 
In addition, some nonconvex optimization algorithms were also brought out for QST, for instance the one in Ref.~\cite{kyrillidis2018provable} is faster and more accurate as compared to previous approaches. 
One notices that all these studies concern only the optimization over quantum state space, with the consideration over quantum measurement space being rarely mentioned.

In fact, various important and meaningful problems related to quantum measurements exist in convex optimization, including for example, searching the Bell parameters in Bell-test experiments \cite{smania2018avoiding}, optimizing the correlation of quantum measurements under different measurement settings \cite{becker2011templates,kleinmann2016quantum,kleinmann2017proposed,hu2018observation}, and maximizing the likelihood functions in quantum measurement tomography.
Meanwhile, characterization of quantum measurements forms the basis for quantum state tomography \cite{cramer2010efficient,torlai2018neural,gross2010quantum} and quantum process tomography \cite{mohseni2008quantum,altepeter2003ancilla,o2004quantum}. Therefore, convex optimization over the quantum measurement space stands as an independent yet important problem in quantum information theory. 
However, the space of quantum measurements is much more complex as compared to the quantum state space since it is possible to produce an infinite variety of different measurement outcomes as long as the probabilities for these outcomes sum to one. Recently, Ref.~\cite{cattaneo2022semidefinite} proposed a method to optimize over the measurement space based on SDP, but it fails to solve complex tasks due to the intrinsic problem with SDP.
Worst of all, nonconvex functions \cite{zhang2020experimental} easily appear in the space of quantum measurements. 
Unlike convex functions, local optima might be found during the process of optimization. 
Hence, nonconvex optimization is regarded as more difficult than convex optimization.
In this work, we propose two reliable algorithms for optimizing arbitrary functions over the space of quantum measurements by combining the so-called Gilbert's algorithm for convex optimization with the direct-gradient (DG) algorithm as well as the accelerated projected gradient (APG) algorithm. With extensive applications, we demonstrate the efficacy of our algorithms with both convex and
nonconvex functions.

This work is organized as follows. In Sec.~\ref{Sec:FunctionOpt}, we propose two reliable algorithms for optimizing over quantum measurement space by combining Gilbert's algorithm with the DG and APG algorithms respectively.
The universality of our method is demonstrated by several examples with both convex and nonconvex functions in Sec.~\ref{Sec:Applications}.
The last Sec.~\ref{Sec:Summary} is the summary.

\section{Function optimization}\label{Sec:FunctionOpt}
In the quantum state space $\mathcal{Q}$, an arbitrary state $\rho$ should satisfy the conditions
\begin{eqnarray}
    \rho&\geq& 0\,,\label{eq:6}\\
    \tr(\rho)&=&1\,.\label{eq:7}
\end{eqnarray}
Given a smaller convex subset ${\mathcal{C} \in \mathcal{Q}}$, Gilbert's algorithm can be used to approximately find the closest state ${\rho^\mathcal{C}\in\mathcal{C}}$ with respect to $\rho$ \cite{shang2018convex}. 
In general, for an arbitrary matrix $M$ in the matrix space $\mathcal{M}$, we employ Gilbert's algorithm to search for the closest quantum state ${\rho^\mathcal{Q}\in\mathcal{Q}}$ with respect to $M$. 
Throughout this work, let's denote the operation by using Gilbert's algorithm as
\begin{equation}
  \rho^{\mathcal{Q}} \equiv \mathcal{S} \bigl(M\bigr)\,.
\end{equation}

Given experimental data, it is critical to identify the measurement settings that are most compatible with the data.
Here, we consider the quantum measurement space $\varOmega$ as all the positive operator-valued measures (POVMs).
A quantum measurement device is characterized by a set of operators $\bigl\{\Pi_l\bigr\}$, which have to satisfy two constraints
\begin{eqnarray}
    \Pi_l&\geq& 0\,,\label{eq:constraints 1}\\
    \sum_{l = 1}^{L}\Pi_l &=&\mathbb{I}\,,\label{eq:constraints 2}
\end{eqnarray}
where $L$ is the total number of operators in the set.
Denote a function $\mathcal{F}\bigl[\bigl\{\Pi _l\bigr\} \bigr]$ defined over the quantum measurement space $\varOmega$. We assume that $\mathcal{F}\bigl[\bigl\{\Pi _l\bigr\} \bigr]$ is differentiable with the gradient $\nabla \mathcal{F}\bigl[\bigl\{\Pi _l\bigr\} \bigr]  \equiv G\bigl[\bigl\{\Pi _l\bigr\} \bigr]$. 
The objective is to optimize $\mathcal{F}\bigl[\bigl\{\Pi_l\bigr\} \bigr]$ over the entire quantum measurement space, and we have 
\begin{subequations}
\begin{eqnarray}\label{eq:2}
\textrm{optimize}&\quad&\,\,\mathcal{F}\bigl[\bigl\{\Pi _l\bigr\} \bigr]\,,\\
\textrm{s.t.}&\quad&\,\,\bigl\{\Pi_l\bigr\} \in \varOmega\,.\label{measurement space}
\end{eqnarray}
\end{subequations}
A simple gradient method is very likely to take $\bigl\{\Pi_l\bigr\}$ outside of the quantum measurement space, for this we employ Gilbert's algorithm to guarantee the condition in Eq.~\eqref{eq:constraints 1}.
In addition, we rewrite the POVM as $\bigl\{ \Pi_l \bigr\} =\bigl\{ \Pi_1,\Pi_2,\dots, \Pi_{L-1}, \mathbb{I}-\sum_{l = 1}^{L-1}\Pi_l\bigr\}$ to satisfy the condition in Eq.~\eqref{eq:constraints 2}. Then, the structure of optimization proceeds as follows.

Taking the to-be-minimized objective function as an example, for the $(k+1)$th iteration, first update the $(L-1)$ elements foremost of the measurement operators with the DG scheme to get
\begin{equation}\label{eq:8}
\begin{split}
   \Pi_{l,k+1}\ &=\Pi_{l,k} -\epsilon G\bigl(\Pi _{l,k}\bigr)\\
    &\equiv DG\Bigl[\Pi_{l,k} ,G\bigl( \Pi _{l,k} \bigr),\epsilon\Bigr].
\end{split}
\end{equation}
Here, $\epsilon$ represents the step size of the update which can be any positive value, and $k$ is the number of iterations.
Second, normalize the measurement operators $\Pi_{l,k+1}$ as density matrices $\rho_{l,k+1}$, such that 
\begin{equation}
  \rho_{l,k+1}=\frac{\Pi_{l,k+1}}{\tr({\Pi_{l,k+1}})}\,,
\end{equation}
which could be nonphysical.
Third, use Gilbert's algorithm to project $\rho_{l,k+1}$ back to the quantum state space $\mathcal{Q}$, i.e., $\rho_{l,k+1} \rightarrow \rho_{l,k+1}^\mathcal{Q}=\mathcal{S} (\rho_{l,k+1})$. 
Finally, reconstruct the physical measurement operators as 
 \begin{equation}\label{eq:tt}
\bigl\{\Pi_{l,k+1}^{\mathcal{\varOmega}}=\rho^\mathcal{Q}_{l,k+1} t_{l,k+1}\bigr\}_{l=1}^{L-1}\,,
 \end{equation}
\begin{equation}\label{povm_last}
  \begin{split}
    \Pi^{\mathcal{\varOmega} }_{L,k+1}&=\mathbb{I}-\sum_{l = 1}^{L-1} \Pi^{\mathcal{\varOmega}}_{l,k+1}\,,
  \end{split}
\end{equation}
where the parameter $t_l$ is obtained by fixing the obtained $\rho_{l,k+1}^\mathcal{Q}$ to get $\bigl\{ t_{l,k+1}\bigr\}_{l=1}^{L-1}=\text{argmin}\ \mathcal{F}\bigl[\bigl\{ t_{l,k+1}\bigr\}_{l=1}^{L-1}\bigr]$.
Here, to ensure that the first $(L-1)$ measurement operators satisfy condition Eq.~\eqref{eq:constraints 1},
only $t_{l,k+1}\geq 0$ is required since $\rho^\mathcal{Q}_{l,k+1}\geq 0$ is guaranteed by using Gilbert's algorithm.
Meanwhile, in order to ensure the last element of the new POVM satisfying the condition in Eq.~\eqref{eq:constraints 1}, let
\begin{equation}\label{povm_last constrain}
    \Pi^{\mathcal{\varOmega} }_{L,k+1}=\mathbb{I} -\sum_{l = 1}^{L-1} \bigl( \rho^{\mathcal{Q} }_{l,k+1} t_{l,k+1}\bigr) \geq 0\,.
\end{equation}
Hence, we get the new POVM $\bigl\{ \Pi^{\mathcal{\varOmega} }_{k+1,l}\bigr\}$ that satisfies the condition in Eq.~\eqref{measurement space} after each iteration.
Whenever the difference between the values of the adjacent iterations is less than a certain threshold, the iteration stops and the optimal POVM is obtained. 
Otherwise, continue with the iteration and the step size is controlled by a step factor $\beta$.
When ${\cF}_{k}<{\cF}_{k-1}$, the step size is appropriately selected. When ${\cF}_{k}>{\cF}_{k-1}$, it indicates that the step size selection is too large, and the step factor $\beta$ needs to be used to adjust the step size.
See the DG algorithm in Algorithm~\ref{alg:algorithm1}.

However, the DG algorithm owns some disadvantages, such as slow optimization speed and low accuracy. For faster convergence, one can choose the APG algorithm \cite{shang2017superfast,beck2009fast}. The APG algorithm adjusts the direction of the gradient at each step, which improves the convergence speed of the algorithm. In simple terms, the APG algorithm has introduced a companion operator $E_{l,k}=\Pi_{l,k}+\frac{\theta_{k-1}-1}{\theta_k}\bigl(\Pi_{l,k}-\Pi_{l,k-1}\bigr)$, which provides the momentum of the previous step controlled by the parameter $\theta$, to
update the measurement operators $\Pi_{l,k}=E _{l,k-1}-\epsilon\,G\bigl(E _{l,k-1}\bigr)$. See the specific process shown in Algorithm~\ref{alg:algorithm2}. 

\begin{minipage}{.9\linewidth}
\begin{algorithm}[H]
    \SetAlgoLined 
	\caption{{\bf DG algorithm}}\label{alg:algorithm1}
	\KwIn{$\epsilon>0$, $0<\beta<1$, choose any $\bigl\{ \Pi_{l,0} \bigr\} _{l=1}^{L-1} \in \varOmega$, $\mathcal{F}_0=\mathcal{F}\bigl[\bigl\{\Pi_{l,0}\bigr\}\bigr]$.}
	\KwOut{$\bigl\{ \Pi_{l} \bigr\} $.}
	\For{$k = 1,\cdots,$}{
		 \For {$l = 1,\cdots,L-1$}{
Update  $\Pi_{l,k}=\mbox{DG}\bigl[ \Pi_{l,k-1},G\bigl( \Pi _{l,k-1}\bigr),\epsilon\bigr]$. Calculate $\rho_{l,k}$ and \ $ \rho_{l,k}^{\mathcal{Q} }=\mathcal{S} \bigl(\rho_{l,k}\bigr)$.
}     
 Gain $\bigl\{ t_{l,k}\bigr\}_{l=1}^{L-1}=\text{argmin}\ \mathcal{F}_k\ \bigl[\bigl\{ t_{l,k}\bigr\}_{l=1}^{L-1}\bigr] $.
 Calculate $\bigl\{\Pi_{l,k}^{\mathcal{\varOmega}}\bigr\}$, $\mathcal{F}_k=\mathcal{F}\bigl[\bigl\{ \Pi_{l,k}^{\mathcal{\varOmega}}\bigr\}\bigr]$. \\
 Termination criterion!\\
     \If{
   ${\cF}_{k}>{\cF}_{k-1}$} { Reset $\epsilon=\beta\epsilon$, and $ \bigl\{\Pi_{l,k}\bigr\} =\bigl\{ \Pi_{l,k-1}^{\mathcal{\varOmega} }\bigr\} $.\ }
  }
\end{algorithm}
\end{minipage}\\

\begin{minipage}{.9\linewidth}
\begin{algorithm}[H]
    \SetAlgoLined 
	\caption{{\bf APG algorithm}}\label{alg:algorithm2}
	\KwIn{$\epsilon>0$, $0<\beta<1$, choose any $\bigl\{\Pi_{l,0} \bigr\} _{l=1}^{L-1} \in \varOmega$, $\bigl\{E_{l,0}\bigr\}=\bigl\{\Pi_{l,0}\bigr\}$, $\theta_0=1$, and $\mathcal{F}_0=\mathcal{F}\bigl[\bigl\{\Pi_{l,0}\bigr\}\bigr]$.}.
	\KwOut{$\bigl\{ \Pi_{l} \bigr\} $.}
	\For{$k = 1,\cdots,$}{
		 \For {$l = 1,\cdots,L-1$}{
Update  $\Pi_{l,k}=E _{l,k-1}-\epsilon\,G\bigl(E _{l,k-1}\bigr)$. Calculate $ \rho_{l,k}$ and\ $\rho_{l,k}^{\mathcal{Q} }=\mathcal{S} \bigl(\rho_{l,k}\bigr)$.
}     
Gain $\bigl\{ t_{l,k}\bigr\}_{l=1}^{L-1}=\text{argmin}\ \mathcal{F}_k\ \bigl[\bigl\{ t_{l,k}\bigr\}_{l=1}^{L-1}\bigr]$.
Calculate $\bigl\{\Pi_{l,k}^{\mathcal{\varOmega}}\bigr\}$, $\mathcal{F}_k=\mathcal{F}\bigl[\bigl\{ \Pi_{l,k}^{\mathcal{\varOmega}}\bigr\}\bigr]$. \\
Termination criterion!\\
\eIf {${\cF}_{k}>{\cF}_{k-1}$} { Reset $\epsilon=\beta\epsilon$, and $\bigl\{ \Pi_{l,k}\bigr\} =\bigl\{ \Pi_{l,k-1}^{\mathcal{\varOmega} }\bigr\} $.\  $\bigl\{E_{l,k}\bigr\}=\bigl\{\Pi_{l,k}\bigr\}$, and $\theta_k=1$.}{Set $\theta_k=\tfrac{1}{2}{\Bigl(1+\sqrt{1+4\theta_{k-1}^2}\Bigr)}$;\\
 Update $\bigl\{E_{l,k}\bigr\}=\bigl\{\Pi_{l,k}+\frac{\theta_{k-1}-1}{\theta_k}\bigl(\Pi_{l,k}-\Pi_{l,k-1}\bigr)\bigr\}$.}
}
\end{algorithm}
\end{minipage}

\section{Applications}\label{Sec:Applications}
In this section, we demonstrate the efficacy of our algorithms by optimizing arbitrary convex as well as nonconvex functions over the space of quantum measurements.

\subsection{Convex functions}
In quantum measurement tomography \cite{natarajan2013quantum,d2004quantum,coldenstrodt2009proposed}, a set of known probe states $\rho_m$ is measured to provide the information needed to reconstruct an unknown POVM $\bigl\{\Pi_l\bigr\}$. The probability that the device would respond to the quantum state $\rho_m$ by producing the outcome $\Pi_l$ is given by
\begin{equation}\label{eq:1} 
p_{lm}=\tr\bigl(\rho_m\Pi_l\bigr)\,.  
\end{equation}
Typically, the linear inversion method \cite{fano1957description} can be used to get the ideal POVM, but nonphysical results are likely to be obtained. Then, the maximum likelihood estimation (MLE) \cite{fiuravsek2001maximum} is proposed to reconstruct the POVM that satisfies all the conditions. However, MLE fails to return any meaningful results when the target POVM is of low rank, which is quite typical especially in higher-dimensional spaces. These problems can be avoided by using our algorithms.

To estimate the operators $\bigl\{\Pi_l\bigr\}$, we maximize the likelihood function
\begin{equation}
  \mathcal{L} \bigl[\bigl\{\Pi_l\bigr\}\bigr]= \prod _{l=1}^{L} \prod_{m=1}^{M}
  \Bigl[\tr\bigl(\rho_m\Pi_l\bigr)\Bigr]^{f_{lm}}\,,
\end{equation}
where $M$ is the number of different input states $\rho_m$, and
\begin{equation}
    f_{lm}=\frac{n_{lm}}{n}\,,
\end{equation}
with $n_{lm}$ denoting the number of $l$th outcome when measuring the $m$th state $\rho_m$, and $n$ representing the total number of measured input states. 
One can see that $\mathcal{L}\bigl [\bigl\{\Pi_l\bigr\}\bigr]$ is not strictly concave, while the log-likelihood 
$\ln \mathcal{L}\bigl[\bigl\{\Pi_l\bigr\}\bigr]$ is. Here, we minimize the negative log-likelihood function $\mathcal{F} \bigl[\bigl\{\Pi_l\bigr\}\bigr]=-\ln \mathcal{L}\bigl[\bigl\{\Pi_l\bigr\}\bigr]$ with
\begin{equation}
  \begin{split}
    \ln \mathcal{L} \bigl[\bigl\{\Pi_l\bigr\}\bigr]=\sum_{l = 1}^{L} \sum_{m = 1}^{M} f_{lm}\ln{p_{lm}}\,.
  \end{split}
\end{equation}
To satisfy the condition in Eq.~\eqref{eq:constraints 2}, rewrite the objective function as
\begin{equation} \label{likelihood function}
    \ln \mathcal{L} \bigl[\bigl\{\Pi_l\bigr\}\bigr]=\sum_{l = 1}^{L-1}\sum_{m = 1}^{M} f_{lm} \ln\bigl[\tr\bigl(\rho_m\Pi_l\bigr)\bigr] +\sum_{m = 1}^{M}  f_{Lm} \ln\biggl\{\tr\biggl[\rho_m\biggl(\mathbb{I} -\sum_{l = 1}^{L-1}\Pi_l \biggr)\biggr]\biggr\}.
\end{equation}
The gradient of $\ln \mathcal{L} \bigl[\bigl\{\Pi_l\bigr\}\bigr]$ with respect to  $\Pi_l$ is
\begin{equation}\label{eq:gradient}
    \begin{split}
        \nabla\ln \mathcal{L}\bigl(\Pi_l\bigr)=&\sum_{m = 1}^{M}\biggl[\biggl(\frac{f_{lm}}{p_{lm}}-\frac{f_{lm}}{1-\sum_{l = 1}^{L-1}p_{lm}}  \biggr)\rho_m\biggr].
      \end{split}
\end{equation}

For numerical simulations, we mainly consider Pauli measurements which are the most commonly-used measurements in quantum information processing.
Then the cases of one qubit, one qutrit, two qubits, and two qutrits are used for the experimental setup respectively.
Specifically, the setups of these four scenarios are described below.

\subsubsection{One qubit}
For one qubit, we take the eigenstates of $\sigma _z$ and the superposition states $-{\frac{1}{\sqrt{2}}}\Bigl(\ket{0_z}\pm \ket{1_z}\Bigr)$ and ${\frac{1}{\sqrt{2}}}\Bigl(\ket{0_z}\pm i\ket{1_z}\Bigr)$ as the input states. In the measurement setup, we select the projection of the spin along the $x$ axis, i.e.,
\begin{equation}
  \Pi_1=\ket{0_x}\bra{0_x}\,;\quad
  \Pi_2=\ket{1_x}\bra{1_x}\,.   
\end{equation}

\subsubsection{One qutrit}
For one qutrit, we use $12$ different input states: 
three eigenstates of $\sigma _z$, $\ket{-1_z}$, $\ket{0_z}$ and $\ket{1_z}$, and nine superposition states ${\frac{1}{\sqrt{2}} }\Bigl(\ket{-1_z}+e^{i\psi _j}\ket{0_z}\Bigr)$, ${\frac{1}{\sqrt{2}} }\Bigl(\ket{0_z}+e^{i\psi _j}\ket{1_z}\Bigr)$ and
${\frac{1}{\sqrt{2}} }\Bigl(\ket{-1_z}+e^{i\varPsi _j}\ket{1_z}\Bigr)$,
where $j=1,2,3$; and $\psi_1=0$, $\psi_2=\frac{\pi}{2}$, and $\psi _3=\pi$. 
The device measures the projection of the spin along the $x$ axis, and the POVM are projectors
\begin{equation}
     \Pi_1=\ket{-1_x}\bra{-1_x}\,;\quad \Pi_2=\ket{0_x}\bra{0_x}\,;\quad \Pi_3=\ket{1_x}\bra{1_x} \,.
\end{equation}

\subsubsection{Two qubits}
In the case of two qubits, we take the tensor products of the
 four eigenstates of two Pauli-$Z$ operators $\ket{0_z0_z}$, $\ket{1_z1_z}$, $\ket{0_z1_z}$, $\ket{1_z0_z}$ and the superposition states ${\frac{1}{\sqrt{2}} }\Bigl(\ket{0_z0_z}+e^{i\psi _j}\ket{0_z1_z}\Bigr)$, 
${\frac{1}{\sqrt{2}} }\Bigl(\ket{0_z0_z}+e^{i\psi _j}\ket{1_z0_z}\Bigr)$, 
${\frac{1}{\sqrt{2}} }\Bigl(\ket{0_z0_z}+e^{i\psi _j}\ket{1_z1_z}\Bigr)$, ${\frac{1}{\sqrt{2}} }\Bigl(\ket{0_z1_z}+e^{i\psi _j}\ket{1_z0_z}\Bigr)$, 
${\frac{1}{\sqrt{2}} }\Bigl(\ket{0_z1_z}+e^{i\psi _j}\ket{1_z1_z}\Bigr)$,
${\frac{1}{\sqrt{2}} }\Bigl(\ket{1_z0_z}+e^{i\psi _j}\ket{1_z1_z}\Bigr)$ as the probe states, where $j=1,2,3$; $\psi_1=0$, $\psi_2=\frac{\pi}{2}$, and $\psi _3=\pi$. Then, we choose the following POVM for the experimental simulation
\begin{equation}
\begin{split}
      &\Pi_1=\ket{0_x0_x}\bra{0_x0_x}\,;\quad\ \Pi_2=\ket{0_x1_x}\bra{0_x1_x}\,;\\
     &\Pi_3=\ket{1_x0_x}\bra{1_x0_x}\,;\quad\  \Pi_4=\ket{0_x0_x}\bra{0_x0_x} \,.\\
\end{split}
\end{equation}

\subsubsection{Two qutrits}
Finally, for the case of two qutrits, we perform a numerical simulation of the Stern-Gerlach apparatus measuring two particles with spin-$1$.
We assume $45$ different input states: $\ket{1_z-1_z}$,
$\ket{-1_z0_z}$, $\ket{-1_z1_z}$, $\ket{0_z-1_z}$, $\ket{0_z0_z}$, $\ket{0_z1_z}$, $\ket{1_z0_z}$, $\ket{1_z1_z}$, $\ket{-1_z-1_z}$,
and $36$ superposition states.
In the simulation, the device measures the projection of the spin along the $x$ 
axis, and the POVM are projectors
\begin{equation}
\begin{split}
    \Pi_1&=\ket{0_x1_x}\bra{0_x1_x}\,;\quad \Pi_2=\ket{0_x-1_x}\bra{0_x-1_x}\,;\\
     \Pi_3&=\ket{1_x0_x}\bra{1_x0_x}\,;\quad \Pi_4 =\ket{1_x-1_x}\bra{1_x-1_x}\,;\\  
     \Pi_5&=\ket{0_x0_x}\bra{0_x0_x}\,; \quad\Pi_6=\ket{-1_x0_x}\bra{-1_x0_x}\,;\\
     \Pi_7&=\ket{1_x1_x}\bra{1_x1_x}\,; \quad\Pi_8=\ket{-1_x1_x}\bra{-1_x1_x}\,;\\ 
      \Pi_9&=\ket{-1_x-1_x}\bra{-1_x-1_x}\,.\\ 
\end{split}
\end{equation}

\begin{figure}[h]
\centering
\subfigure[one qubit]{
\includegraphics[width=.4\columnwidth]{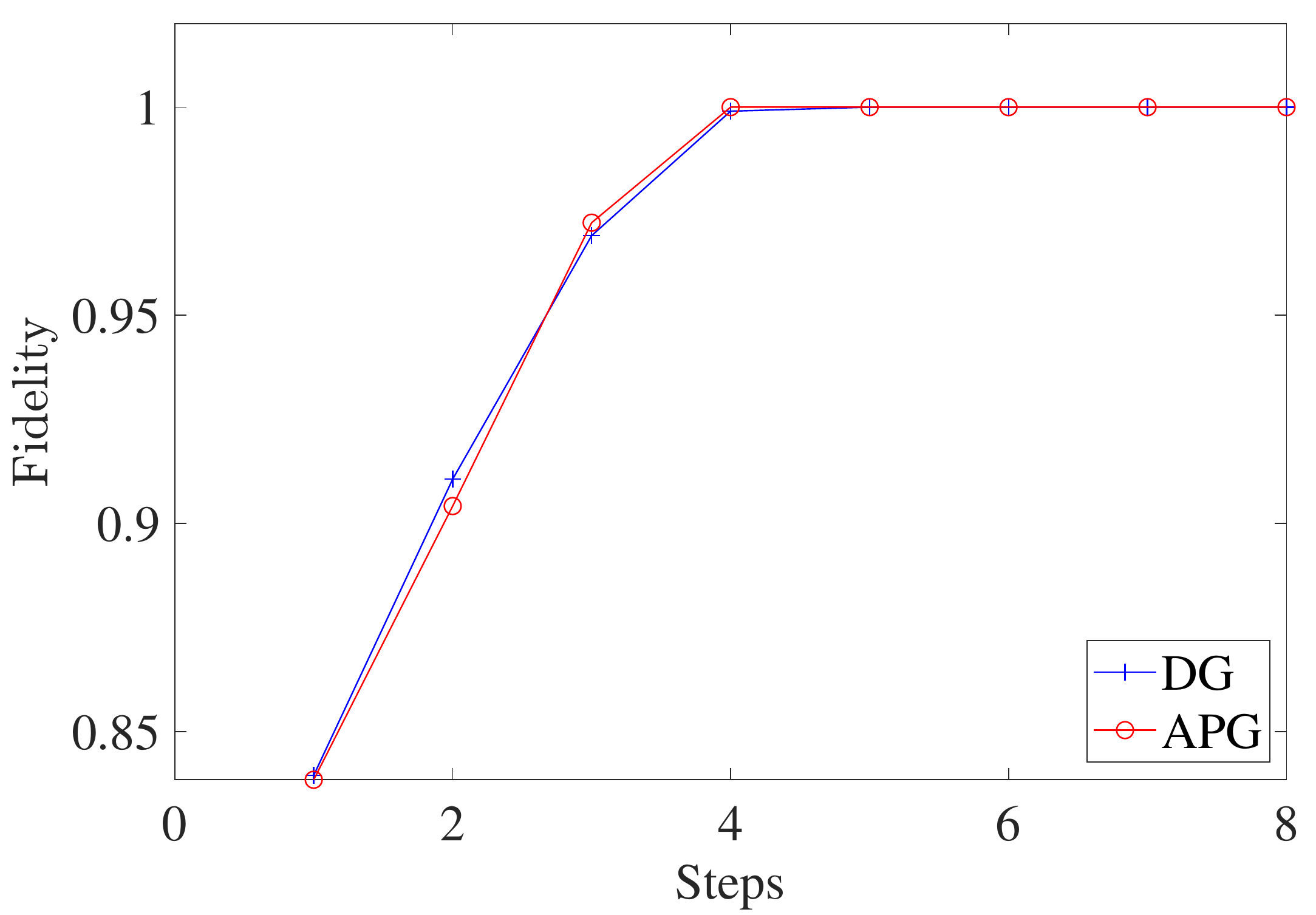}
}
\quad
\subfigure[one qutrit]{
\includegraphics[width=.4\columnwidth]{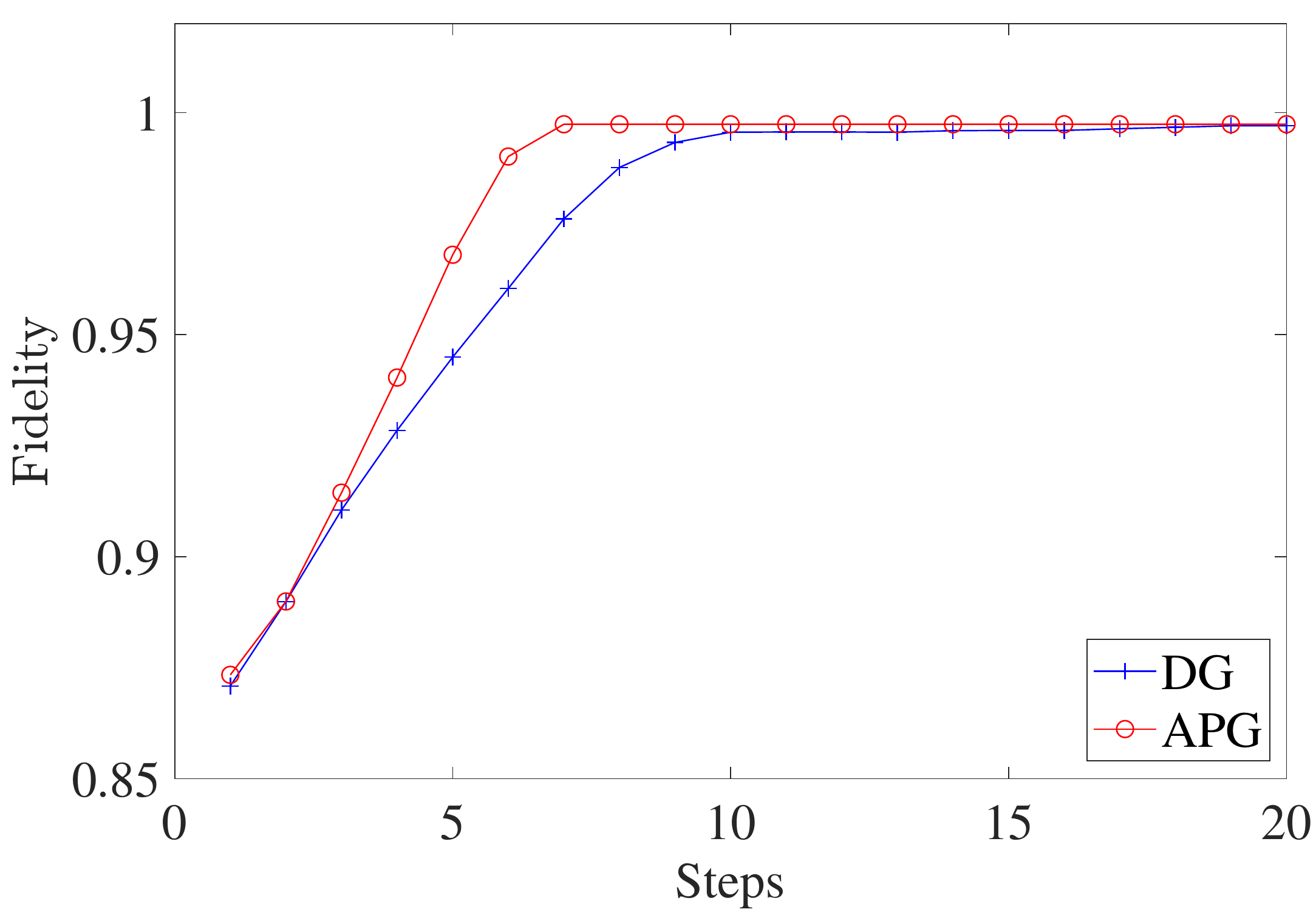}
}
\quad
\subfigure[two qubits]{
\includegraphics[width=.4\columnwidth]{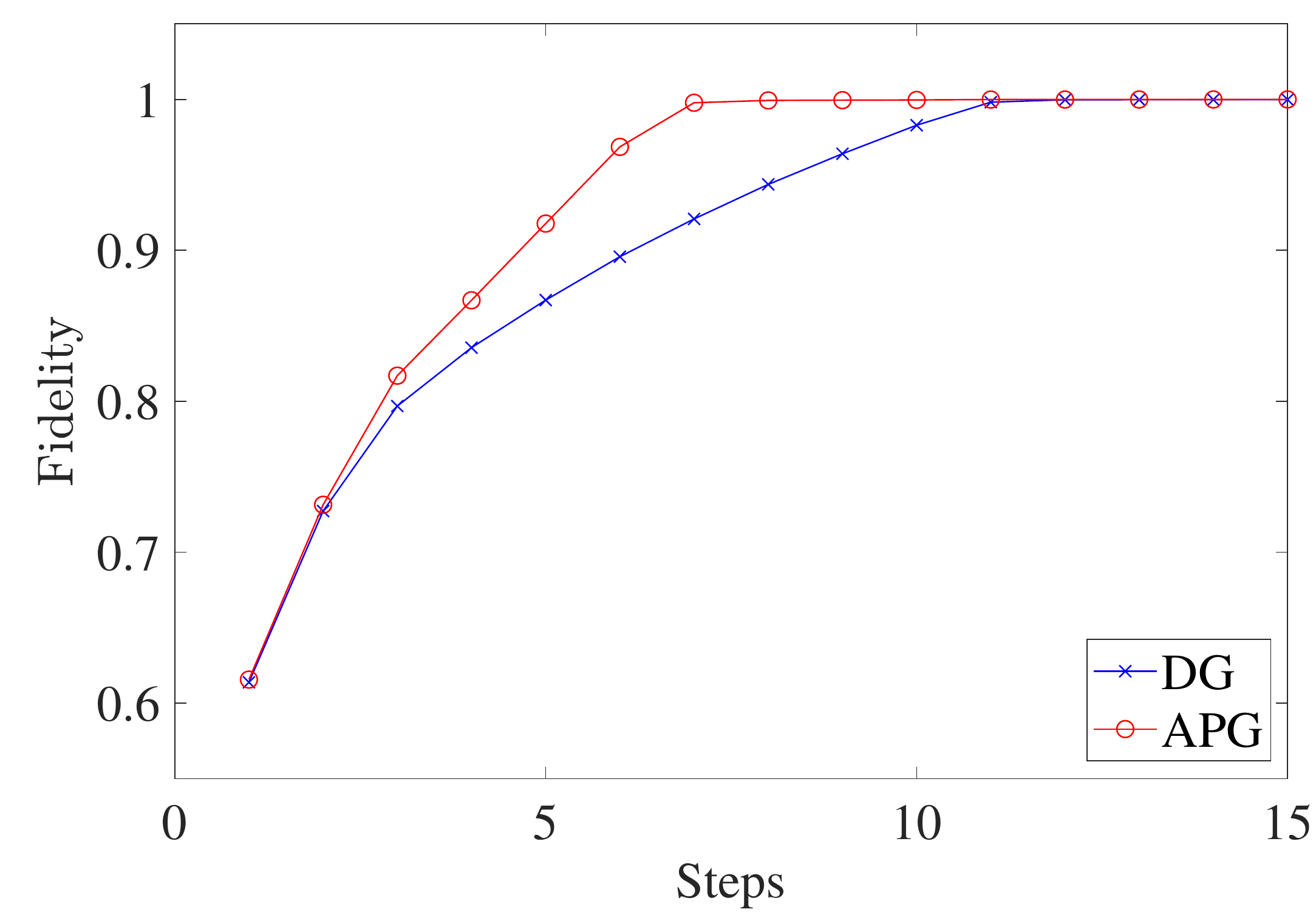}
}
\quad
\subfigure[two qutrits]{
\includegraphics[width=.4\columnwidth]{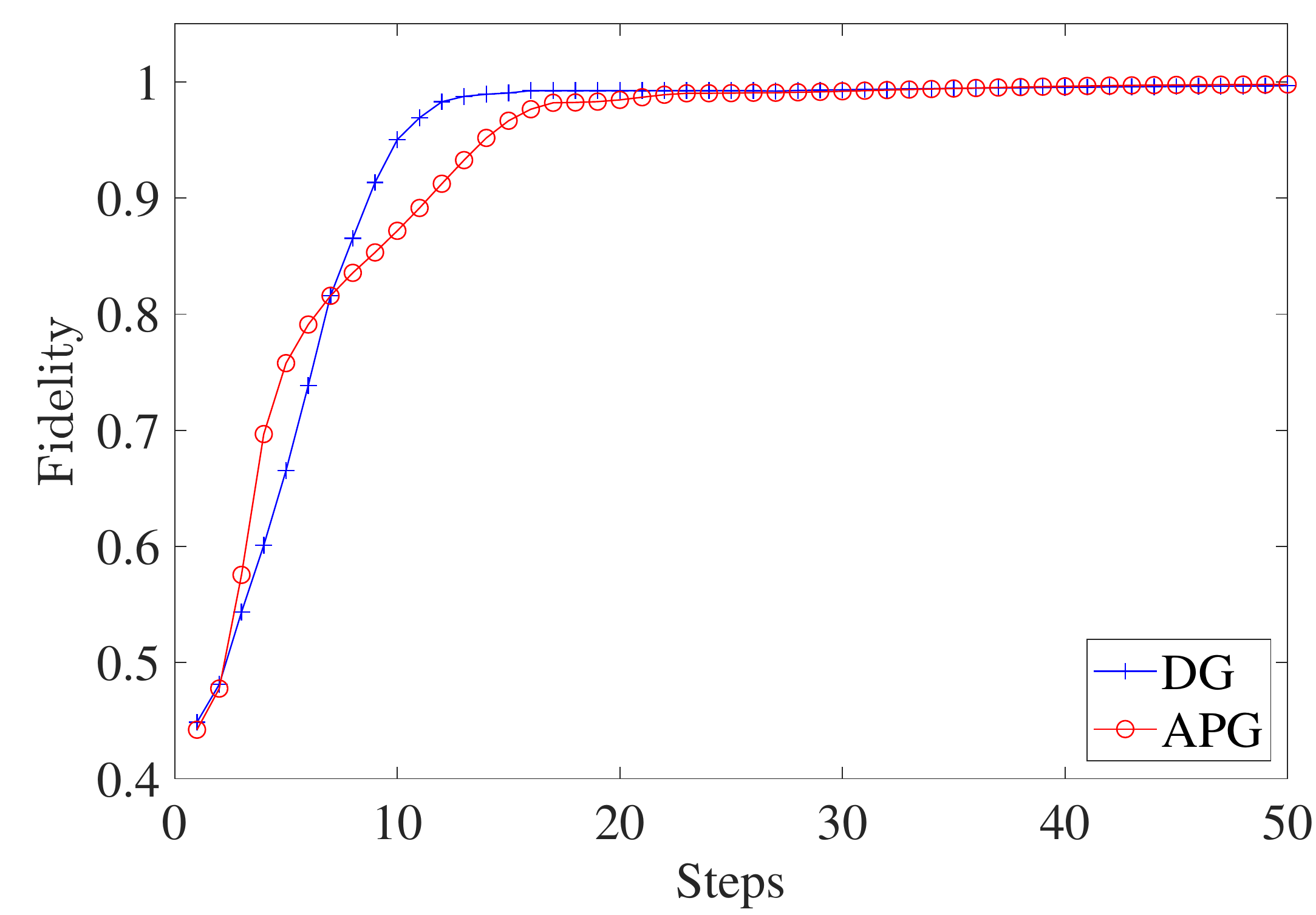}
}
\caption{For different cases, the two algorithms are compared to reconstruct the overall fidelity of the measurements. 
The number of measurements used in each simulation for each probe state is $300$, $10^5$, $10^5$, and $5\times10^5$ respectively.
For most cases, the APG algorithm converges faster than the DG algorithm.
}\label{compare}
\end{figure}

\begin{figure}[h]
\centering
\subfigure[one qubit]{
\includegraphics[width=.4\columnwidth
]{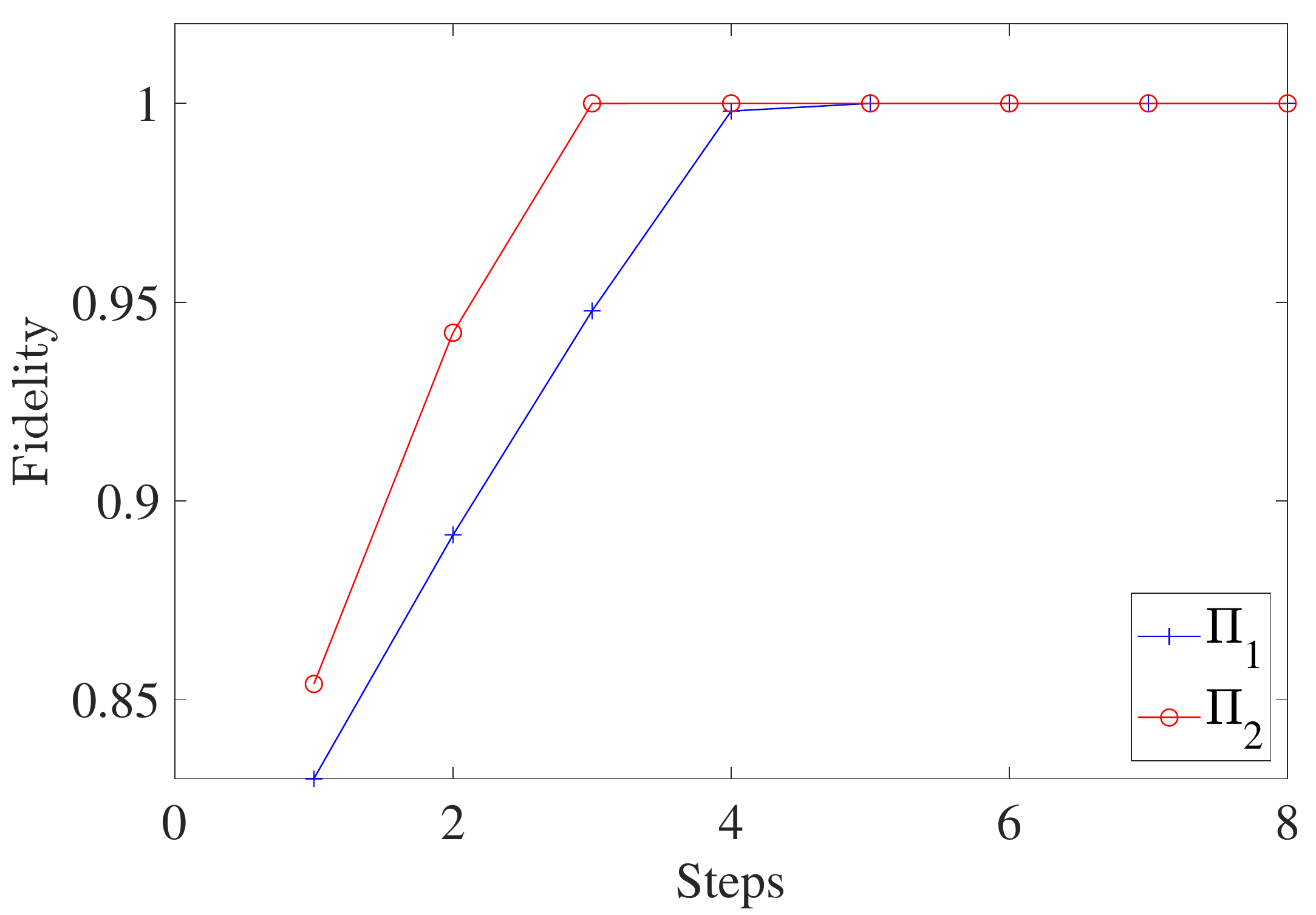}
} 
\quad
\subfigure[one qutrit]{
\includegraphics[width=.4\columnwidth]{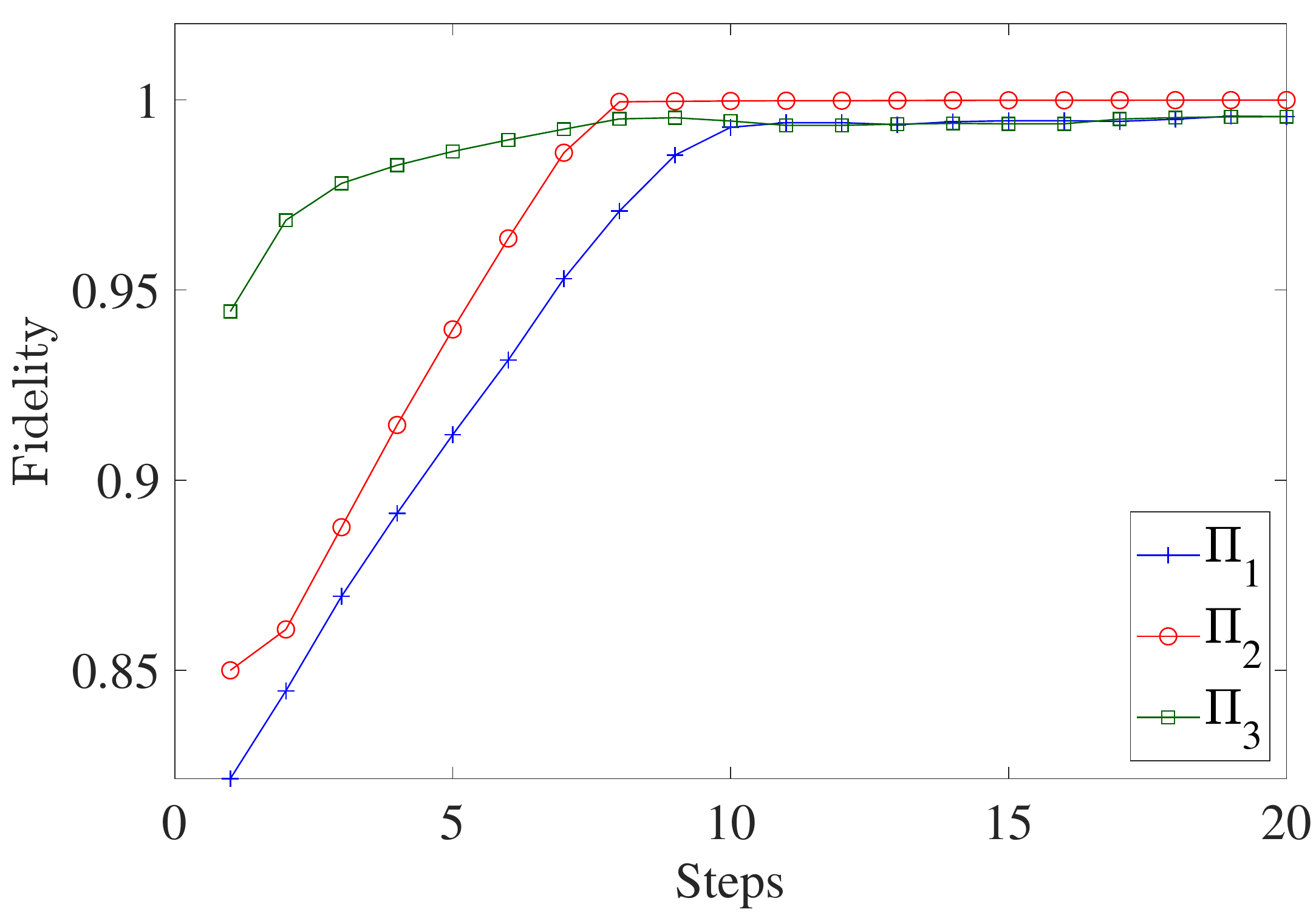}
}
\quad
\subfigure[two qubits]{
\includegraphics[width=.4\columnwidth]{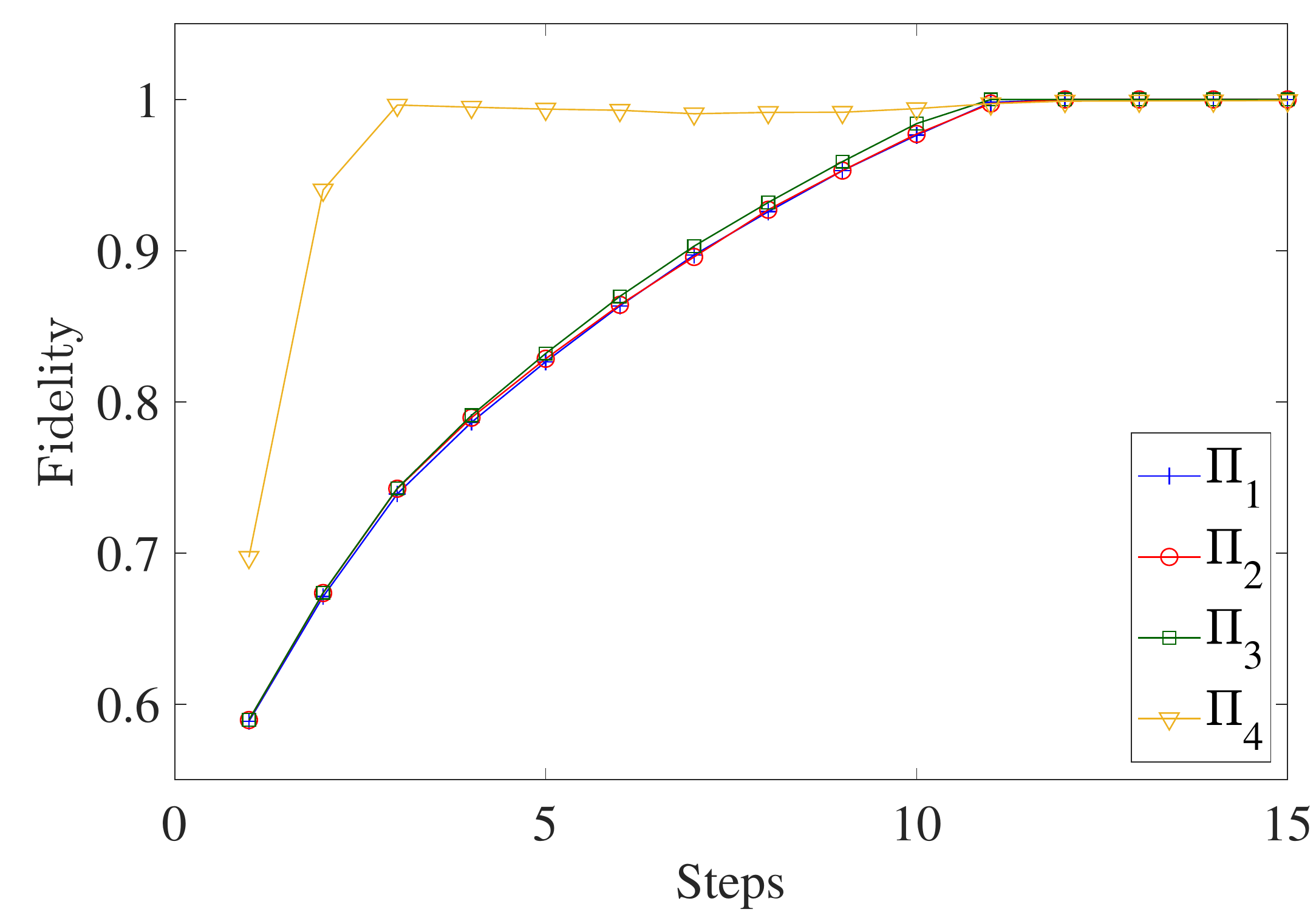}
}
\quad
\subfigure[two qutrits]{
\includegraphics[width=.4\columnwidth]{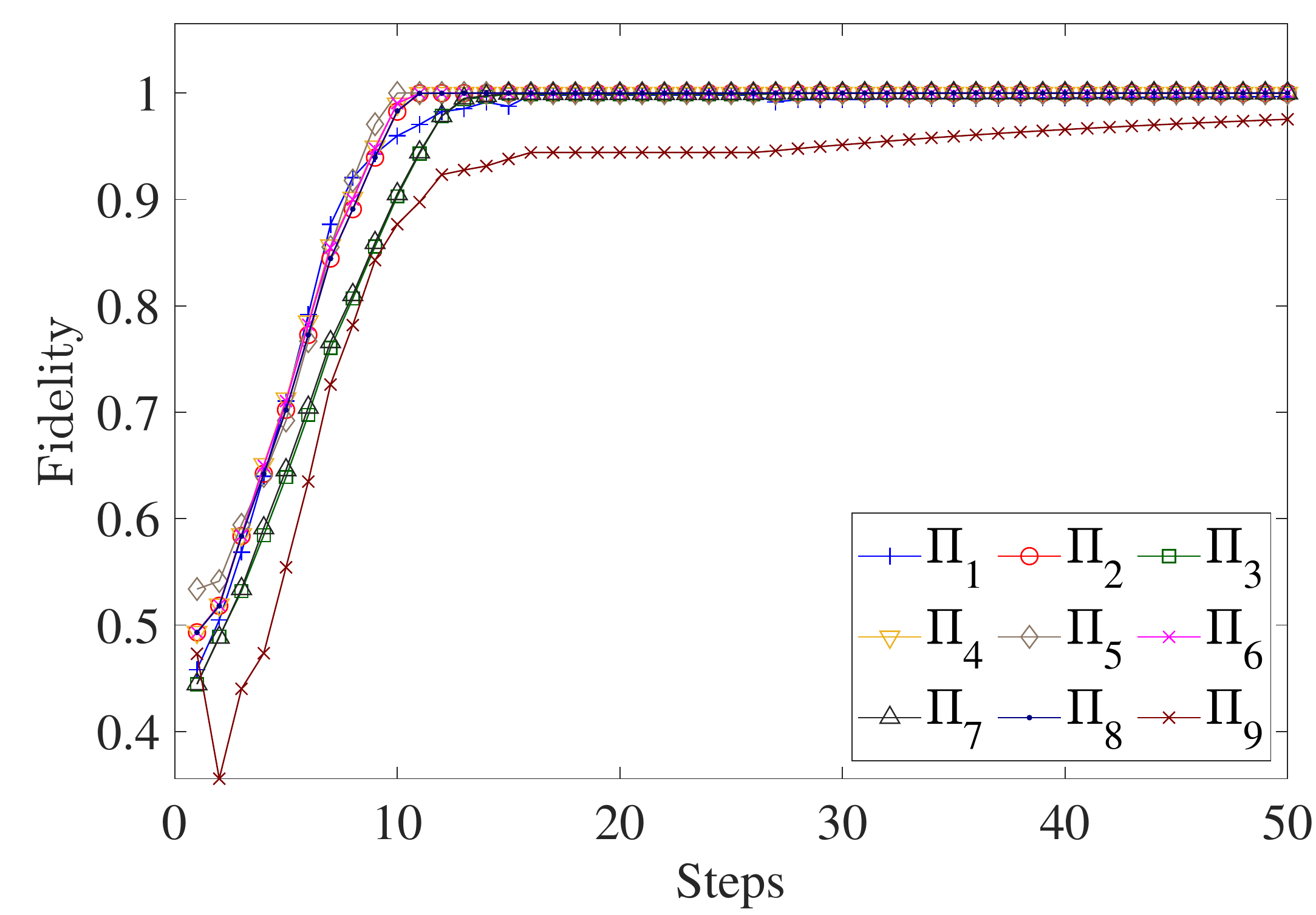}
} 
\caption{For different cases of the quantum measurement tomography, fidelities of the measurements obtained by the DG algorithm vary with the number of iteration steps.
In general, the fidelity of each POVM element saturates to the maximum very quickly.}\label{all element dg}
\end{figure}

\begin{figure}[h]
\centering
\subfigure[one qubit]{
\includegraphics[width=.4\columnwidth]{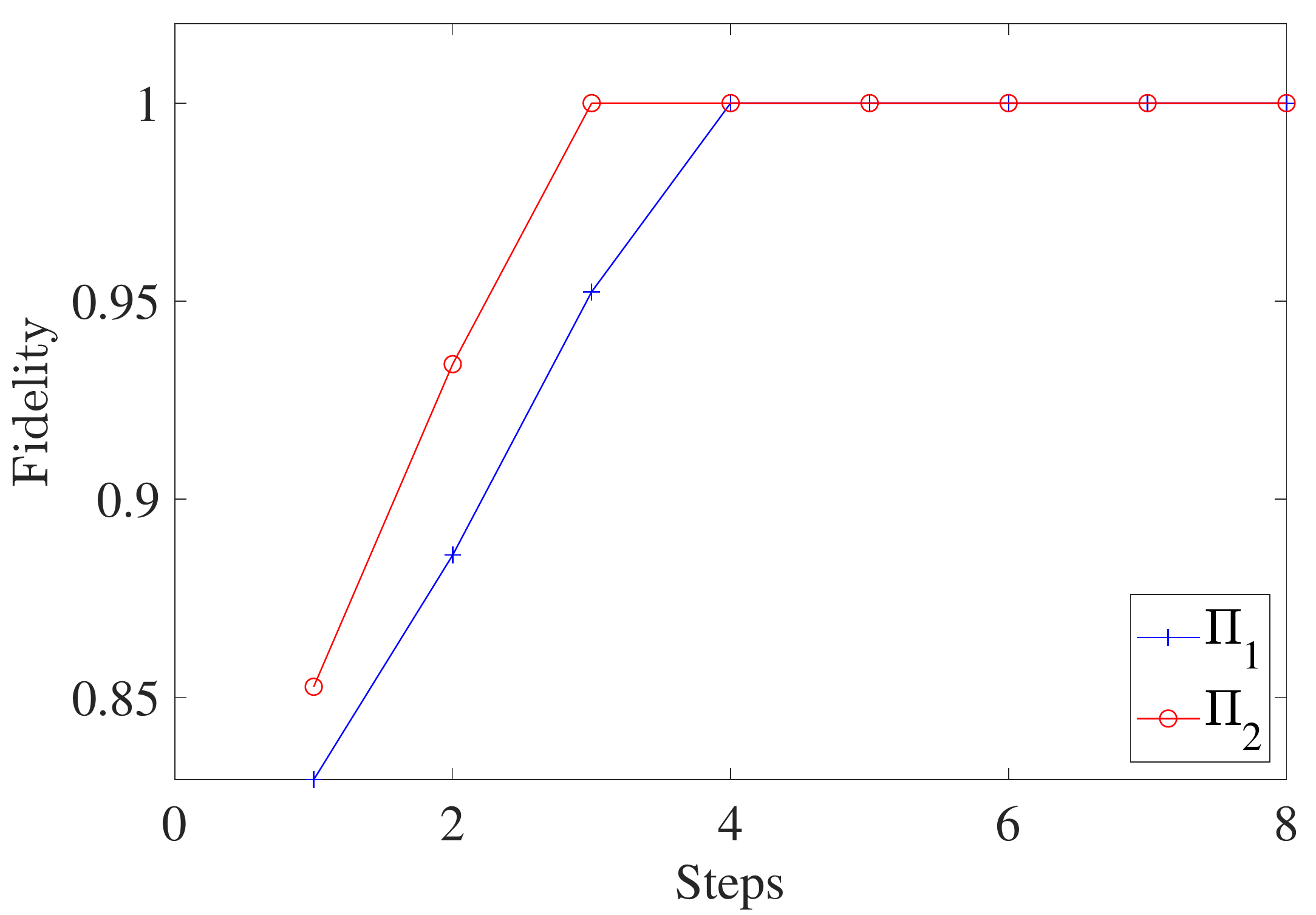}
} 
\quad
\subfigure[one qutrit]{
\includegraphics[width=.4\columnwidth]{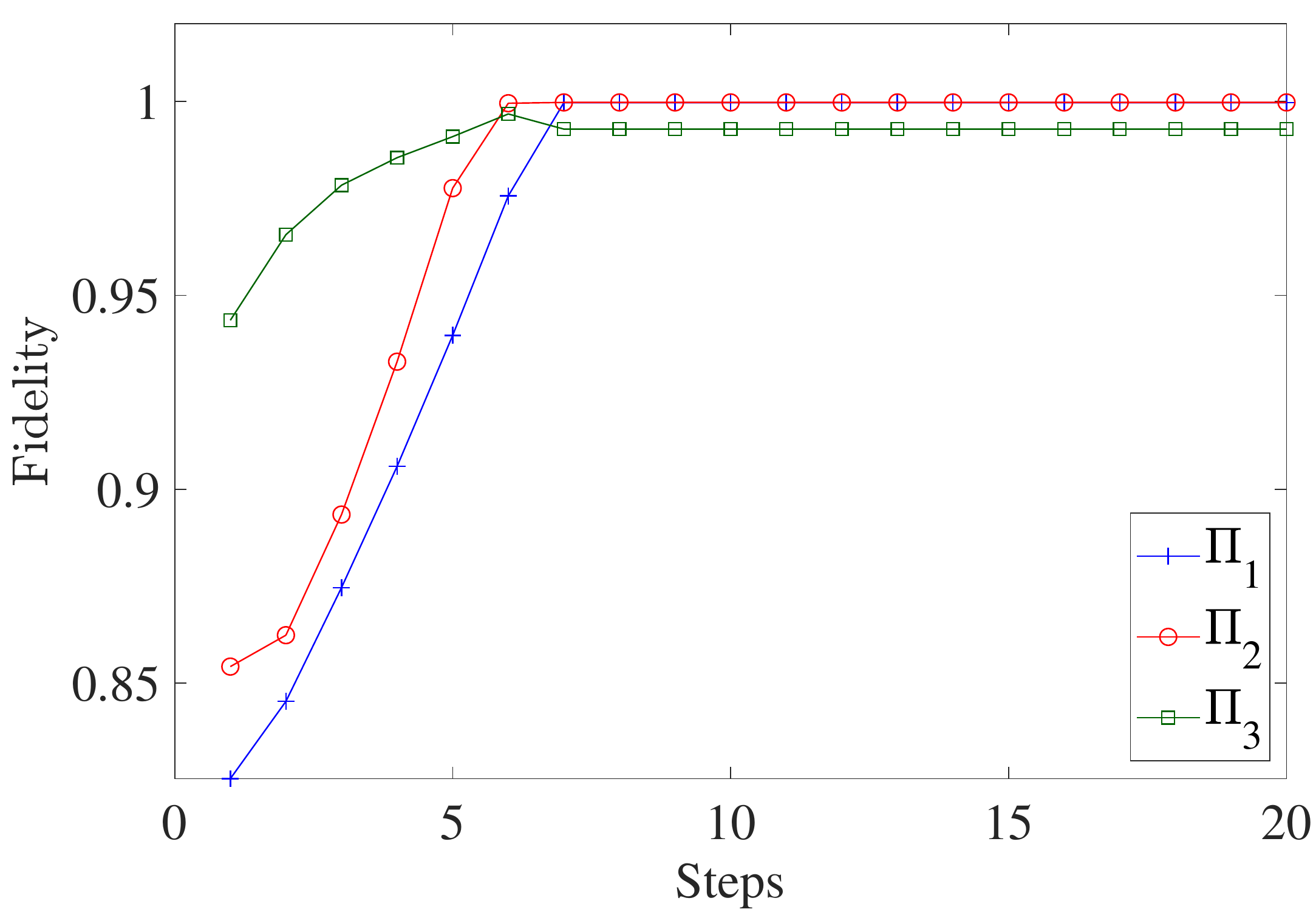}
}
\quad
\subfigure[two qubits]{
\includegraphics[width=.4\columnwidth]{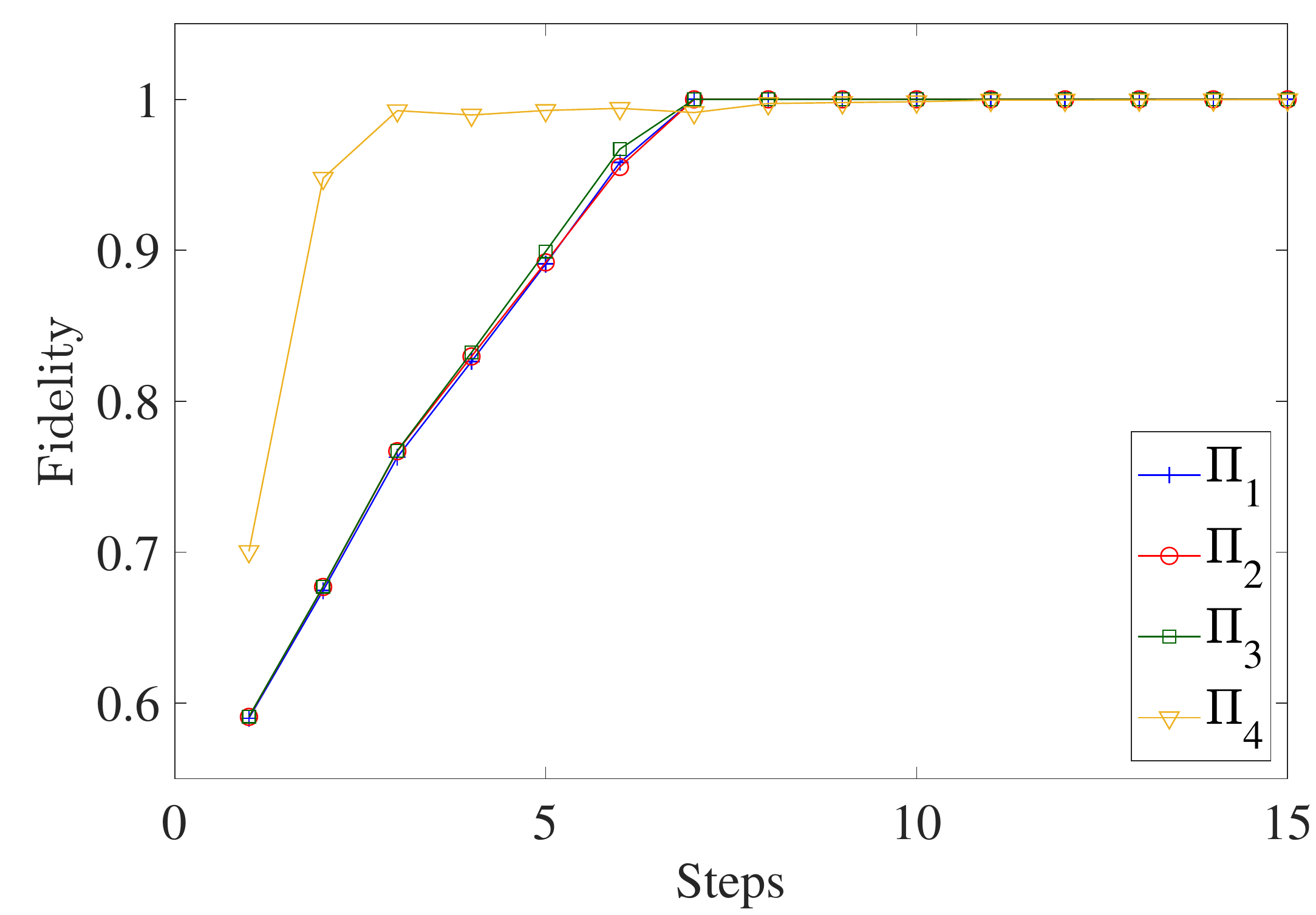}
} 
\quad
\subfigure[two qutrits]{
\includegraphics[width=.4\columnwidth]{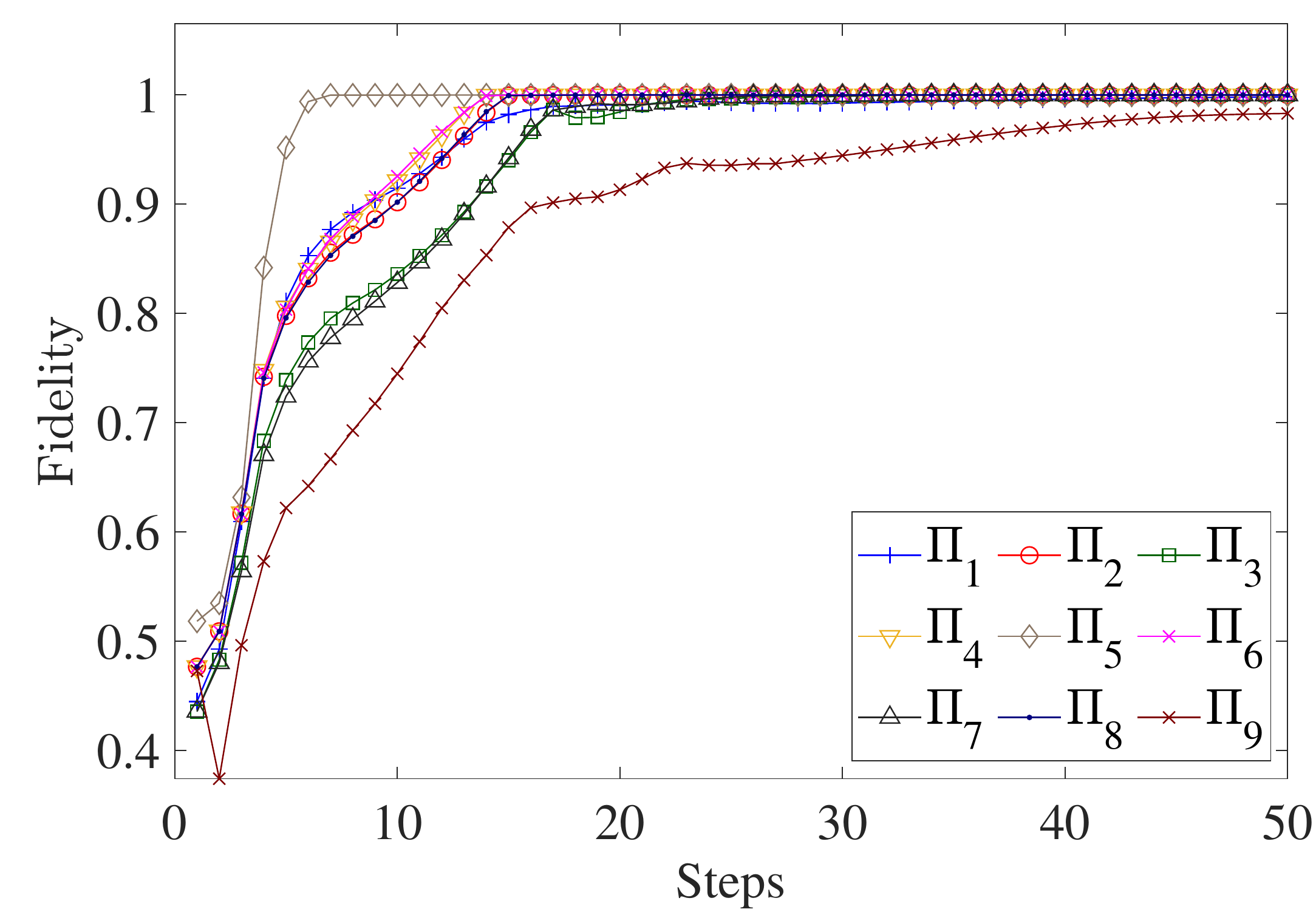}
} 
\caption{For different cases of the quantum measurement tomography, fidelities of the measurements obtained by the APG algorithm vary with the number of iteration steps. In general, the fidelity of each POVM element saturates to the maximum very quickly.}\label{all element apg}
\end{figure}

For each case of simulation, the number of measurements for each probe state is $300$, $10^5$, $10^5$, and $5\times10^5$ respectively.
Then according to the frequency obtained by the simulated data, we use our algorithm to reconstruct the POVM.
The fidelity between different POVM elements is defined as the
fidelity between the two states $\sigma$ and $\rho$, i.e.,
\begin{equation}
\begin{split}
    F(\sigma, \rho)\coloneqq\biggl(\tr\sqrt{\sqrt{\sigma} \rho \sqrt{\sigma} }\biggr)^{2}= F\biggl(\frac{\Pi_l}{\tr(\Pi_l) } \,, \frac{\Pi_{j}}{\tr(\Pi_{j})}\biggr).
\end{split}
\end{equation} 
In addition, the overall fidelity between two POVMs $\bigl\{\Pi_l\bigr\}_{l=1}^L$ and $\bigl\{\Pi_{j}\bigr\}_{j=1}^L$ on an
$d$-dimensional Hilbert space is defined by
\begin{equation}
    F(\Pi_l, \Pi_j)\coloneqq\left[\sum_{l = 1}^{L}  w_{l}\sqrt{F\Bigl(\frac{\Pi_l}{\tr(\Pi_l) } , \frac{\Pi_{j}}{\tr(\Pi_{j})}\Bigr)}\right]^{2} \!,
\end{equation}
with $w_l = \frac{\sqrt{\tr(\Pi_{l})\tr(\Pi_{j})}}{d} $ \cite{hou2018deterministic}. 
The overall fidelities of the reconstructed POVMs are shown in Fig. \ref{compare}. 
Figures~\ref{all element dg} and \ref{all element apg} present the variations of fidelity of the POVM elements reconstructed using DG algorithm and APG algorithm with respect to the number of iteration steps in different cases. We can see that these two algorithms are almost identical in accuracy, and the fidelities of the measurement operators are close to 1.
Generally speaking, the APG algorithm converges faster than the DG algorithm. 
In addition, one notices that the fidelity of the last element in some of the simulations is not always increasing, which is a result of the constraint that we set in Eq.~\eqref{povm_last constrain}.


\subsection{Nonconvex functions}
Quantum detector self-characterization (QDSC) tomography is another 
method for characterizing quantum measurements.
Unlike quantum measurement tomography, this method does not require to know the specific form of the input probe states, but directly optimizes the cost function based on the measurement statistic $\bm{f_{m}}$ to reconstruct the measurements. For POVM with $L$ outcomes detected by $m$ states, a data set of the measurement statistic $f_{lm}$ is obtained.  We write the distribution of the data for each state as a vector
\begin{equation}
\bm{f_{m}}=\left(
\begin{array}{c}
    f_{1m}\\
    f_{2m}\\
    \vdots\\
    f_{Lm}
\end{array}
\right)\,.
\end{equation}

For the one qubit case, define $N_{i,l}=\bm{b}_i^T\bm{b}_l$ and write the POVM as
\begin{equation} \label{bloch povm}
    \Pi_l=a_lI+\bm{b}_l\cdot \bm{\sigma}
\end{equation}
under the Bloch representation, where $i$ and $l$ represent the number of rows $i$ and columns $l$ of the matrix $N$, $\bm{a}=(a_{1}\cdots a_{L})^{T}$, $\bm{b}_l=(b_{l,x},b_{l,y},b_{l,z})$, $\bm{\sigma}=(\sigma_x,\sigma_y,\sigma_z)$, $1\leq  i,l \leq L $. The matrix $N$ and vector $\bm{a}$ can be represented as 
\begin{equation}\label{eq:Q_kl}
\begin{split}
    N_{i,l}&=\bm{b}_i^T\bm{b}_l=\frac{1}{2} \tr(\Pi_i\Pi_l)-\frac{1}{4} \tr(\Pi_i)\tr(\Pi_l)\,,\\
\end{split}
\end{equation}
\begin{equation}\label{eq;constrain a}
a_l=\frac{1}{2} \tr(\Pi_l)\,.
\end{equation}
Then, optimization of the cost function $\mathcal{F}\bigl(N^+,\bm{a}\bigr)$ is given by \cite{zhang2020experimental}
\begin{subequations}
\begin{eqnarray}
\textrm{min}&\quad&\,\,\sum_{m} \biggl[1-\bigl(\bm{f_{m}}-\bm{a}\bigr)^T N^+ \bigl(\bm{f_{m}}-\bm{a}\bigr)\biggr]^2\,,\label{eq:zhang}\\
\textrm{s.t.}&\quad&\,\, a_l^2-N_{l,l}>=0\,,\label{QDSC constrain}
\end{eqnarray}
\end{subequations}
where $N^+$ stands for the Moore-Penrose pseudoinverse of $N$. 
One notices that the objective function is nonconvex.
Optimization of nonconvex functions is difficult as local minima might be found.
Interestingly, we find that our algorithm can also be used to optimize nonconvex functions. Since our algorithm guarantees the conditions for quantum measurements, one only needs to optimize the objective function regardless of the constraint in Eq.~\eqref{QDSC constrain}.

For numerical simulations, we  choose $50$ probe states:
\begin{equation}
    \frac{1}{2}\bigl(\mathbb{I}+\sigma_z\bigr)\,,\quad  \frac{1}{2}\bigl(\mathbb{I}-\sigma_z\bigr)\,,\quad 
    \frac{1}{2}\biggl(\mathbb{I}+\sin{\frac{i\pi}{4}}\cos{\frac{n\pi}{8}}\sigma_x+\sin{\frac{i\pi}{4}}\sin{\frac{n\pi}{8}}\sigma_y+\cos{\frac{i\pi}{4}}\sigma_z\biggr)\,,
\end{equation}
where $i=1,2,\cdots,6$; $n=1,2,\cdots,8$. And we use the two-dimensional SIC POVM as the measurement device, and each state is measured $200$ times. The APG algorithm is used to optimize the objective function.
First, select any set of POVM operators in the measurement space, use Eqs.~\eqref{eq:Q_kl} and \eqref{eq;constrain a} to obtain the initial values $N_k^+$ and $\bm{a}_k$ respectively.
Similarly, we calculate the gradient of the objective function in Eq.~\eqref{eq:zhang}. The gradient of the objective function is given by
\begin{equation}
    \delta \mathcal{F} \bigl(\bm{a}\bigr) =\sum_{m}2\bigl(1-\bm{f_{m}}-\bm{a}\bigr)^TN^+\bigl(\bm{f_{m}}-\bm{a}\bigr) \biggl\{\bigl(N^+\bigr)^T \bm{f_{m}}+N^+\bm{f_{m}}-\bigl[N^++\bigl(N^+\bigr)^T\bigr] \bm{a}\biggr\}\, ,
\end{equation}
\begin{equation}
    \delta \mathcal{F} \bigl(N^+\bigr)=\sum_{m}-2\bigl(1-\bm{f_{m}}-\bm{a}\bigr)^TN^+\\
    \bigl(\bm{f_{m}}-\bm{a}\bigr)^2\bigl(\bm{f_{m}}-\bm{a}\bigr)^T \,.
\end{equation}
The values of $N_{k+1}$ and $\bm{a_{k+1}}$ are obtained by iterating over $N_{k}$ and $\bm{a_{k}}$ using gradient descent, then $\bm{b}_{l,k+1}$ is obtained by decomposing $N_{k+1}$. In the experiment, we specify that the reference frame, i.e., the vector $\bm{b}_{1}$ is parallel to the $z$ direction of the Bloch sphere, and set the $xz$ plane of the Bloch sphere as the plane determined by the vectors $\bm{b}_{1}$ and $\bm{b}_{2}$. This is equivalent to $b_{1,x}=b_{1,y}=b_{2,y}=0$. 
Then, $\bigl\{ \Pi_{l,k+1}\bigr\}_{l=1}^{L-1}$ can be obtained by using Eq.~\eqref{bloch povm}, which is the update for $\bigl\{ \Pi_{l,k}\bigr\}_{l=1}^{L-1}$.

\begin{figure}[t]
    \centering
     \includegraphics[width=.5\columnwidth]{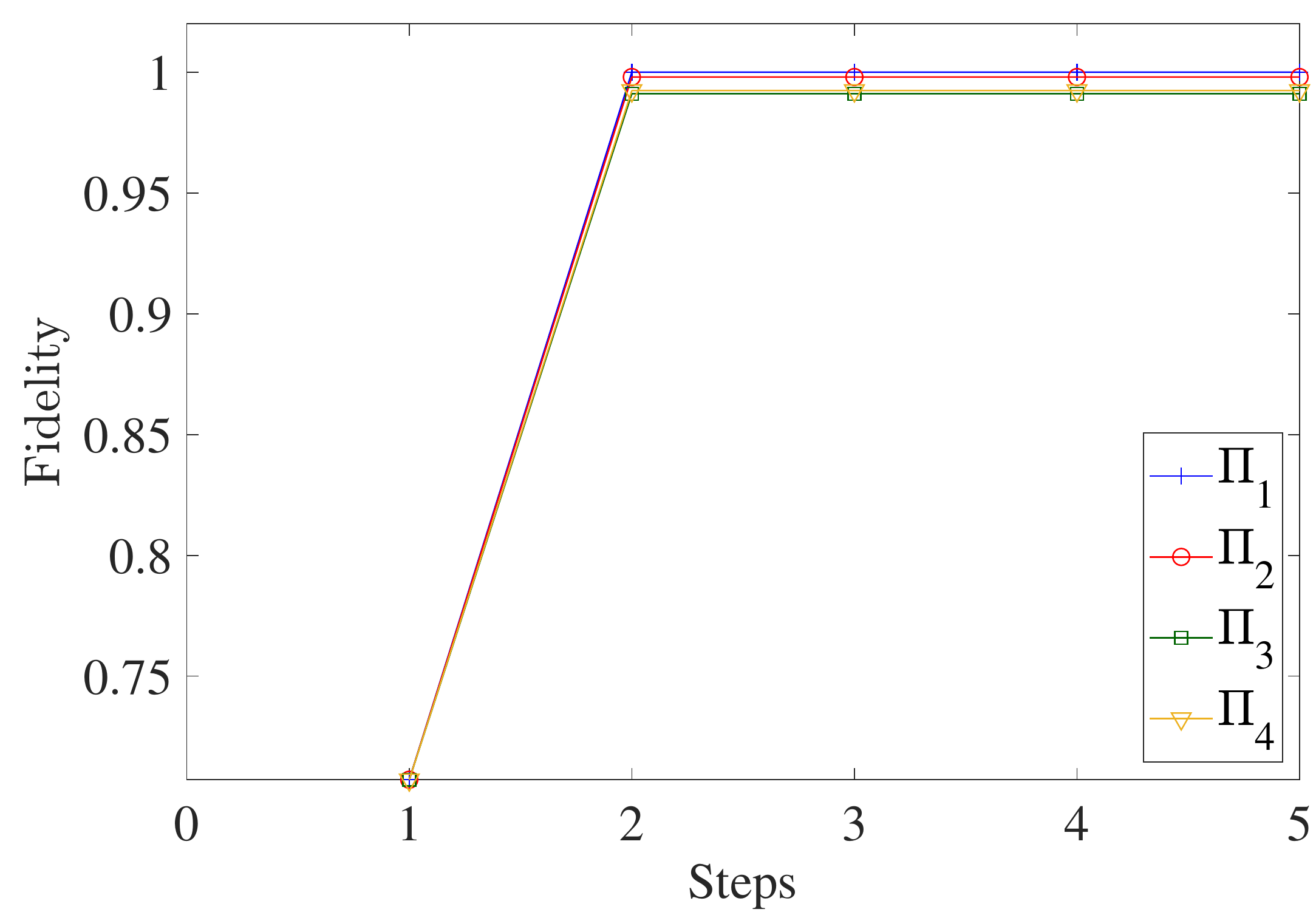}
 
     \caption{In the case of QDSC, the fidelity of each element of the two-dimensional SIC POVM saturates to the maximum by using only 2 steps.}     \label{QDSC_element_fidelity}
\end{figure}

\begin{figure}[t]
    \centering
     \includegraphics[width=.5\columnwidth]{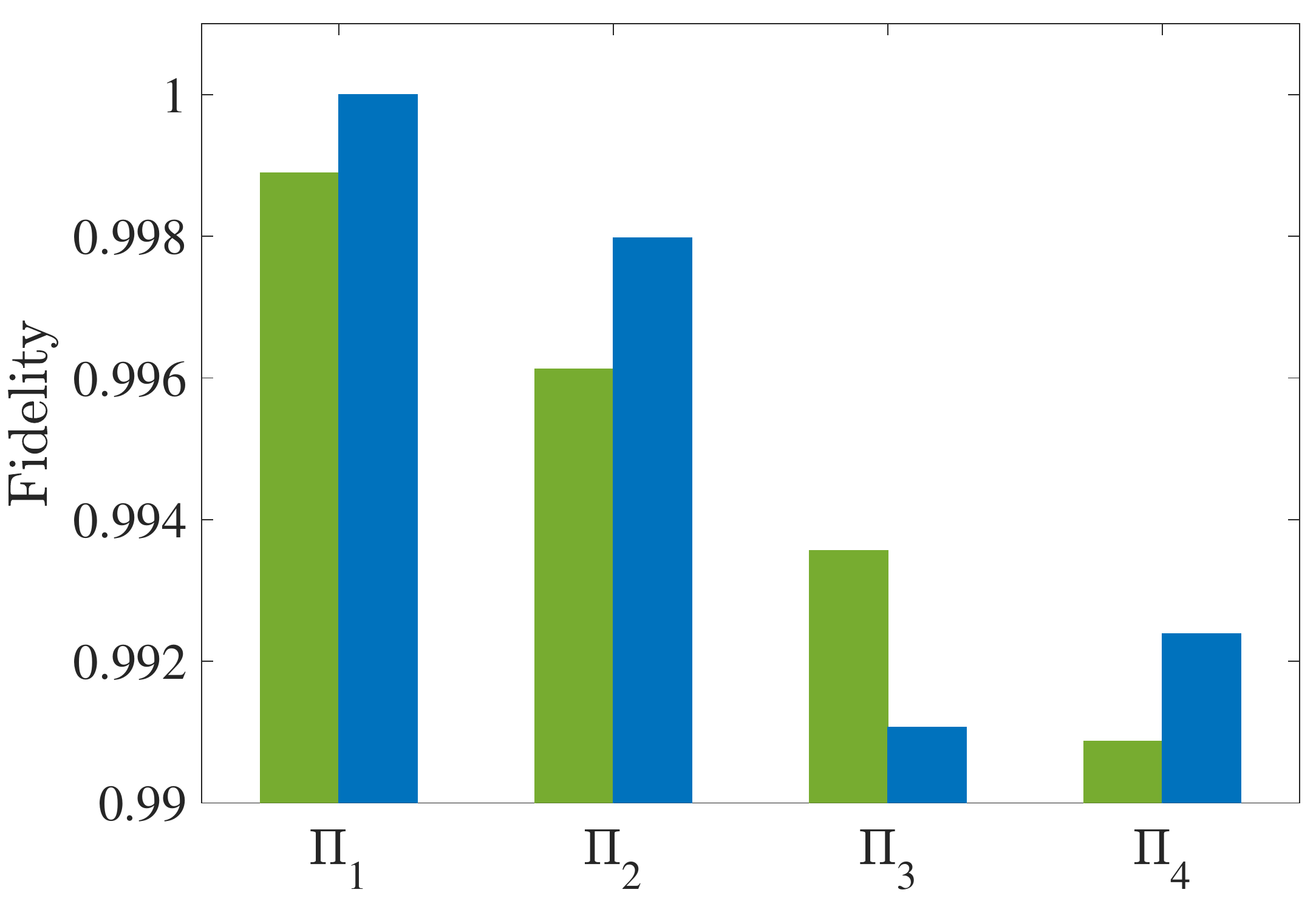}
     \caption{Comparison of the fidelities of the reconstructed quantum measurements between the APG algorithm (blue) and the method in  Ref.~\cite{zhang2020experimental} (green).}
     \label{figure:2}
\end{figure}

The fidelity of each POVM element can approach $1$ in a very small number of iteration steps; see Fig.~\ref{QDSC_element_fidelity}.
Then the fidelities of the measurements are compared with the ones reported in Ref.~\cite{zhang2020experimental}, demonstrating that the performance of our algorithm is slightly better; see Fig.~\ref{figure:2}. 

\section{Summary}\label{Sec:Summary}
We have proposed two reliable algorithms for optimizing arbitrary functions over the quantum measurement space. For demonstration, we have shown several examples on the convex function of quantum measurement tomography with different dimensions as well as nonconvex function of one qubit in quantum detector self-characterization tomography. Surprisingly, our method does not encounter the problem of rank deficiency. Compared with SDP, our method can be easily applied to higher-dimensional cases as well as to optimize nonconvex functions.
Moreover, our method reports better results as compared to previous approaches.
For future work, we will consider the optimization over the joint space of quantum states and quantum measurements, for tasks such as calculating the capacity of quantum channels.


\vspace{6pt}

\authorcontributions{JL performed the numerical calculations.
All authors contributed to the interpretation of the results, preparation and writing of the manuscript.}

\funding{This work has been supported by the National Natural Science Foundation of China (Grants No.~11805010, No.~12175014, and No.~92265115).}

\acknowledgments{We thank Ye-Chao Liu for fruitful discussions.}

\conflictsofinterest{The authors declare no conflicts of interest.}

\reftitle{References}

\end{document}